\documentstyle[aps,pra]{revtex}

\begin{document}
\title{Quantum strategies of quantum measurements}
\author{Chuan-Feng Li,\thanks{%
Electronic address: cfli@ustc.edu.cn} Yong-Sheng Zhang, Yun-Feng Huang, and
Guang-Can Guo\thanks{%
Electronic address: gcguo@ustc.edu.cn}}
\address{Laboratory of Quantum Communication and Quantum Computation and Department\\
of Physics,\\
University of Science and Technology of China, Hefei 230026, People's\\
Republic of China\vspace{0.5in}}
\maketitle

\begin{abstract}
In classical Monty Hall problem, one player can always win with probability $%
2/3$. We generalize the problem to the quantum domain and show that a fair
two-party zero-sum game can be carried out if the other player is permitted
to adopt quantum measurement strategy.\newline

\end{abstract}

Recently, quantum game theory\cite{EWL99,BH00,MEYER99,GVW99,ZHANG00} has
drawn much attentions, which is the combination of applied mathematics and
quantum information theory. It has been shown that quantum strategies can be
more successful than classical ones, based on the principles of quantum
mechanics, such as quantum entanglement\cite{EWL99,BH00} and quantum
superposition\cite{MEYER99,GVW99}. Here we show that quantum measurement,
which has been successful in lots of quantum processes, such as quantum
programming\cite{NIELSEN}, the purification of entanglements\cite{BEN96},
probabilistic teleportation\cite{WANLI} and cloning\cite{DUAN}, can also
contribute to quantum strategies.

A novel quantum game has been introduced by Goldenberg $et$ $al.$\cite{GVW99}%
, where a particle and two boxes are involved. Here, we consider the Monty
Hall problem\cite{MONTY}, a two-party zero-sum game involving a particle and
more boxes.

In the classical case, Alice may put the particle in one of some (such as
three) boxes and Bob picks one box. If he finds the particle in this box, he
wins (suppose one coin), otherwise, he loses one coin. Obviously, Bob will
not agree to this proposal, for he has definite $2/3$ chance to lose. He may
argue that after he chooses one box, Alice should reveal {\it an empty one
from the other two} boxes and then he is provided a chance to choose between
sticking with the original choice and switching to the third box.
Counterintuitively, this puts them in a dilemma situation. Bob can win with
probability $2/3$ by choosing to switch\cite{MONTY,WWW}. The key point is
that when Bob selects a box, he has expected $1/3$ chance to win, which will
not change anymore. Under the condition that Alice reveals an empty box, Bob
may have $2/3$ chance to find the particle in the third box.

Inspired by the previous nice works\cite{EWL99,BH00,MEYER99,GVW99}, we may
ask if Alice can change the situation and play a fair game with Bob provided
with quantum strategies.

A quantum version of Monty Hall problem may be as follows: there are one
quantum particle and three boxes $0,$ $1$ and $2$ (the quantum delineation
is a particle with three eigenstates $\left| 0\right\rangle $, $\left|
1\right\rangle $ and $\left| 2\right\rangle $). Alice puts the particle into
the boxes (maybe in a superposition state). Bob picks one box. In order to
determine the location of the particle, Alice may measure the particles
herself before she reveals an empty box to Bob. At last Bob makes his
decision on sticking or switching.

Obviously, in the classical problem, Alice is forbidden to pick the particle
to other boxes after he places it (otherwise, it becomes an ordinary
gambling with two boxes). It's a natural generalization that in the quantum
case Alice cannot evolve the particle unitarily even if she has an auxiliary
particle, that is (i) single bit rotations (ii) controlled-NOT operations%
\cite{UNIGATE} on the particle are prohibited. While both in classical and
quantum problem, measurement is permitted.

We first consider the quantum superposition and entanglement strategies.
Alice may prepare the particle in a superposition state, such as 
\begin{equation}
\left| \psi \right\rangle _p=\frac 1{\sqrt{3}}\left( \left| 0\right\rangle
+\left| 1\right\rangle +\left| 2\right\rangle \right) .  \label{1}
\end{equation}
After his selection, Bob has $1/3$ chance to win. According to rule (i),
Alice cannot change the probability distribution (quantum measurement
strategy will be discussed below). After Alice points out an empty box (von
Neumann measurement under the basis $\left\{ \left| 0\right\rangle ,\left|
1\right\rangle ,\left| 2\right\rangle \right\} $ is needed), Bob still has $%
2/3$ chance to win by switching. Alice can also entangle the particle with
an auxiliary particle, i.e., the particles are in an entangled state, such
as 
\begin{equation}
\left| \Psi \right\rangle _{p,aux}=\frac 1{\sqrt{3}}\left( \left|
0\right\rangle \left| 0\right\rangle _{aux}+\left| 1\right\rangle \left|
1\right\rangle _{aux}+\left| 2\right\rangle \left| 2\right\rangle
_{aux}\right) .  \label{2}
\end{equation}
According to the rules (i) and (ii), what Alice can do is only local
operations on the auxiliary particle. Obviously, these operations cannot
affect the local probability distribution of the playing particle, which
means the dilemma situation remains unchanged.

Now, we suppose that Alice is allowed to adopt the quantum measurement
strategies. For simplicity, we first assume that Alice put the particle in
one of the Boxes initially. The box Bob chooses is assigned $\left|
0\right\rangle $, and Alice reveals no particle in $\left| 2\right\rangle $.
At this moment, the state of the particle may be described by 
\begin{equation}
\rho _p=\frac 13\left| 0\right\rangle \left\langle 0\right| +\frac 23\left|
1\right\rangle \left\langle 1\right| .  \label{3}
\end{equation}
One property of quantum measurement is that there exists collective
measurement on the Boxes. Because Alice knows exactly which box the particle
is in, his strategy may depend on the choice of Bob. For the cases the
particle in $\left| 0\right\rangle $ and in $\left| 1\right\rangle $,
suppose that the orthogonal bases of the von Neumann measurement Alice
adopts are 
\begin{equation}
\begin{array}{c}
\left| \phi \right\rangle _0=\cos \alpha _0\left| 0\right\rangle +\sin
\alpha _0\left| 1\right\rangle , \\ 
\left| \phi \right\rangle _0^{\bot }=-\sin \alpha _0\left| 0\right\rangle
+\cos \alpha _0\left| 1\right\rangle ,
\end{array}
\label{4}
\end{equation}
and 
\begin{equation}
\begin{array}{c}
\left| \phi \right\rangle _1=\cos \alpha _1\left| 0\right\rangle +\sin
\alpha _1\left| 1\right\rangle , \\ 
\left| \phi \right\rangle _1^{\bot }=-\sin \alpha _1\left| 0\right\rangle
+\cos \alpha _1\left| 1\right\rangle ,
\end{array}
\label{5}
\end{equation}
respectively, where $0\leq \alpha _0,\alpha _1\leq \frac \pi 2$. The density
operator of the particle becomes 
\begin{equation}
\rho _p^{\prime }=\frac 13\left( \cos ^2\alpha _0\left| \phi \right\rangle
_0\left\langle \phi \right| +\sin ^2\alpha _0\left| \phi \right\rangle
_0^{\bot }\left\langle \phi \right| \right) +\frac 23\left( \cos ^2\alpha
_0\left| \phi \right\rangle _1\left\langle \phi \right| +\sin ^2\alpha
_0\left| \phi \right\rangle _1^{\bot }\left\langle \phi \right| \right) .
\label{6}
\end{equation}

Bob's two pure strategies, sticking on $\left| 0\right\rangle $, and
switching to $\left| 1\right\rangle $, are denoted by $N$ and $V$,
respectively, satisfying 
\begin{equation}
N\rho =\left\langle 0\right| \rho \left| 0\right\rangle ,\text{ }V\rho
=\left\langle 1\right| \rho \left| 1\right\rangle .  \label{7}
\end{equation}
We first confine Bob's strategies in a classical mixture region, that is 
\begin{equation}
S_B\left( \eta \right) =\eta N+\left( 1-\eta \right) V,  \label{8}
\end{equation}
where $0\leq \eta \leq 1$. The expected probability for Bob to win is 
\begin{equation}
\begin{array}{c}
P_B=S_B\left( \eta \right) \rho _p^{\prime }=\frac 13\left[ \eta \left( \cos
^4\alpha _0+\sin ^4\alpha _0\right) +2\left( 1-\eta \right) \sin ^2\alpha
_0\cos ^2\alpha _0\right] \\ 
+\frac 23\left[ \left( 1-\eta \right) \left( \cos ^4\alpha _1+\sin ^4\alpha
_1\right) +2\eta \sin ^2\alpha _1\cos ^2\alpha _1\right] .
\end{array}
\label{9}
\end{equation}
The corresponding expected gain for Bob is 
\begin{equation}
G_B=2P_B-1.  \label{10}
\end{equation}
It is easy to find the Nash equilibrium\cite{NASH} of the problem: Alice
adopts a deterministic strategy 
\begin{eqnarray}
\left| \phi \right\rangle _0 &=&\left| \phi \right\rangle _1=\frac 1{\sqrt{2}%
}\left( \left| 0\right\rangle +\left| 1\right\rangle \right) ,  \label{11} \\
\left| \phi \right\rangle _0^{\bot } &=&\left| \phi \right\rangle _0^{\bot }=%
\frac 1{\sqrt{2}}\left( -\left| 0\right\rangle +\left| 1\right\rangle
\right) ,  \nonumber
\end{eqnarray}
while Bob adopts a probabilistic strategy 
\begin{equation}
S_B\left( \frac 12\right) =\frac 12\left( N+V\right) .  \label{12}
\end{equation}
The equilibrium expected winning probability and gain of Bob are 
\begin{equation}
P_B^e=\frac 12,\text{ }G_B^e=0.  \label{13}
\end{equation}
During the deducing of the Nash equilibrium, our discussion is limited in a
simple case, i.e., initially, Alice puts the particle in one of the three
boxes and Bob chooses between sticking and switching at the end. Next, we
simply show that the Nash equilibrium is stable in their whole strategic
spaces: Alice may put the particle in a superposition state, while Bob may
choose the quantum superposition of sticking and switching $R(\beta )$, with 
\begin{equation}
R(\beta )\rho =(\cos \beta \left\langle 0\right| +\sin \beta \left\langle
1\right| )\rho (\cos \beta \left| 0\right\rangle +\sin \beta \left|
1\right\rangle ).  \label{14}
\end{equation}
If Alice adopts the strategy of Equations $(11)$, according to Eq. $(6)$,
the density operator of the particle becomes 
\begin{equation}
\rho _p^{\prime }=I.  \label{15}
\end{equation}
No matter what strategy Bob adopts, pure or mixture of Eq. $(14)$, $G_B\leq
G_B^e=0$ is always satisfied. On the other hand, if Bob adopts the strategy
of Eq. $(12)$, then, 
\begin{equation}
S_B\left( \frac 12\right) \rho =\frac 12\left( N+V\right) \rho =\frac 12%
tr\rho \equiv \frac 12,  \label{16}
\end{equation}
which is independent of Alice's strategy (the density operator of the
particle). Therefore, neither of them can improve his/her expected gain ($0$
in this case) by changing his/her strategy while the other player does not.
That is, the pair of strategies of Eqs. $(11)$ and $(12)$ is a Nash
equilibrium.

Obviously, there is no deterministic Nash equilibrium, the reason lies in
that Alice's strategy has two variable while Bob's pure strategy has one.

The power of the quantum measurement strategy may be embodied more clearly
in the generalized $N$-stage Monty Hall problem, which can be delineated as
follows: Step 1, Alice puts a particle in one of $N$ boxes. Bob picks one
box. Step 2, Alice reveals an empty one from the other $N-1$ boxs and gives
Bob the option of switching to one of the other $N-2$ boxes. Step 3, Alice
reveals another empty boxes, etc. After $N$ stages, the gambling finishes.
The best classical strategy of Bob is: stick until the last choice, then
switch. The optimal expected winning probability is $\frac{N-1}N$.
Obviously, Bob will definitely win, provided $N\rightarrow \infty $.
However, if Alice is allowed to adopt quantum measurement strategy, it is
easy to verify that there still exists a Nash equilibrium: Alice waits until
the last step to adopt the strategy of Eqs. (11), while Bob sticks until the
last choice to adopt the strategy of Eq. (12). Their expected gains are both 
$0$, which means Alice can still play a fair game with Bob by adopting
quantum measurement strategy. The main reason is that in quantum
measurement, there exists collective measurement, while in classical
measurement, there's only orthogonal measurement.

In conclusion, we have generalized the Monty Hall problem to the quantum
domain. It is shown that a fair two-party zero-sum game can be carried out
if a player is permitted to adopt quantum measurement strategy, while in
classical situation, the other player can always win with high probability.

We thank Dr. Derek Abbott for calling our attention to the present problem
and helpful discussions. This work was supported by the National Natural
Science Foundation of China.\

\end{document}